\documentclass[sigconf]{acmart}
\usepackage{url}
\usepackage{multirow}
\usepackage{balance}
\usepackage{rotating}
\usepackage{booktabs}
\usepackage{hyphsubst}
\usepackage{graphicx}
\usepackage{amsmath}
\graphicspath{{images/}}
\usepackage{times}
\usepackage{hhline}
\usepackage[caption = false]{subfig}

\usepackage{url}
\usepackage{rotating}
\usepackage{amsmath}
\usepackage{cancel}
\usepackage{amssymb}
\usepackage{enumitem}
\usepackage{pifont}

\usepackage{multirow}
\usepackage{array}
\usepackage{placeins}

\usepackage{listings}

\usepackage{color}

\newcommand\YAMLcolonstyle{\color{orange}\mdseries}
\newcommand\YAMLkeystyle{\color{black}\bfseries}
\newcommand\YAMLvaluestyle{\color{blue}\mdseries}

\makeatletter

\newcommand\language@yaml{yaml}

\expandafter\expandafter\expandafter\lstdefinelanguage
\expandafter{\language@yaml}
{
  keywords={true,false,null,y,n},
  keywordstyle=\color{blue},
  basicstyle=\YAMLkeystyle,                                 
  sensitive=false,
  comment=[l]{\#},
  morecomment=[s]{/*}{*/},
  commentstyle=\color{red}\ttfamily,
  stringstyle=\YAMLvaluestyle\ttfamily,
  moredelim=[l][\color{orange}]{\&},
  moredelim=[l][\color{magenta}]{*},
  moredelim=**[il][\YAMLcolonstyle{:}\YAMLvaluestyle]{:},   
  morestring=[b]',
  morestring=[b]",
  literate =    {---}{{\ProcessThreeDashes}}3
                {>}{{\textcolor{orange}\textgreater}}1     
                {|}{{\textcolor{orange}\textbar}}1 
                {\ -\ }{{\mdseries\ -\ }}3,
}

\lst@AddToHook{EveryLine}{\ifx\lst@language\language@yaml\YAMLkeystyle\fi}
\makeatother

\newcommand\ProcessThreeDashes{\llap{\color{cyan}\mdseries-{-}-}}

\definecolor{pblue}{rgb}{0.13,0.13,1}
\definecolor{pgreen}{rgb}{0,0.5,0}
\definecolor{pred}{rgb}{0.9,0,0}
\definecolor{pgrey}{rgb}{0.46,0.45,0.48}

\usepackage{rotating}
\usepackage{booktabs}
\usepackage{rotating}
\usepackage{hhline}
\usepackage{hyphsubst}
\usepackage{soul}

\newlength{\Oldarrayrulewidth}

\acmConference[EYRE'18]{1st International Workshop on EntitY REtrieval co-located with CIKM 2018}{October 2018}{Lingotto, Turin,  Italy} 
\acmYear{2018}
\copyrightyear{2018}

\begin{document}

\title[Neighborhood Troubles]{Neighborhood Troubles: On the Value of User Pre-Filtering To Speed Up and Enhance Recommendations}

\author{Emanuel Lacic}
\affiliation{%
  \institution{Know-Center GmbH}
  \city{Graz, Austria} 
}
\email{elacic@know-center.at}

\author{Dominik Kowald}
\affiliation{%
  \institution{Know-Center GmbH}
  \city{Graz, Austria} 
}
\email{dkowald@know-center.at}

\author{Elisabeth Lex}
\affiliation{%
  \institution{Graz University of Technology}
  \city{Graz, Austria} 
}
\email{elisabeth.lex@tugraz.at}

\begin{abstract}
In this paper, we present work-in-progress on applying user pre-filtering to speed up and enhance recommendations based on Collaborative Filtering. We propose to pre-filter users in order to extract a smaller set of candidate neighbors, who exhibit a high number of overlapping entities and to compute the final user similarities based on this set. To realize this, we exploit features of the high-performance search engine Apache Solr and integrate them into a scalable recommender system. We have evaluated our approach on a dataset gathered from Foursquare and our evaluation results suggest that our proposed user pre-filtering step can help to achieve both a better runtime performance as well as an increase in overall recommendation accuracy.
\end{abstract}

\keywords{
User Pre-Filtering;
Entity Filtering;
Recommender Systems;
Real-Time Recommendations;
Apache Solr;
Collaborative Filtering
}

\maketitle

\section{Introduction}
\label{sec:introduction}
In the past decade, there has been a vast amount of research in the field of recommender systems, mostly focusing on developing novel recommendation algorithms \cite{Shani2011} and improving recommender accuracy \cite{rana2015study}. Thus, many well known methods are available, such as Content-based Filtering \cite{Balabanovic:1997:FCC:245108.245124}, 
Collaborative Filtering \cite{sarwar2001item} or Matrix Factorization \cite{koren2009matrix}, which all have their unique strengths and weaknesses.
With the arrival of the big data era, recommender systems are nowadays not only expected to analyze a lot of data, but also to handle frequent streams of new data. 
Traditional recommender systems usually analyze the data offline and update the generated model in regular time intervals.  However, choices made by users depend on factors that are susceptible to change anytime and to re-train such models tends to be a time-consuming task (especially when the data is sparse \cite{Salakhutdinov2008}).

As such, the attention of the recommender systems' research community has recently shifted towards recommendation systems that process streaming data online and recommend entities in near real-time.
For example, recent work from Huang et al. \cite{Huang2015} presented TencentRec, a real-time recommender system that is based on Apache Storm. Specifically, they tackle item-based Collaborative Filtering and handle the data sparsity problem by recommending most popular entities from the user's demographic group. 
Another scalable item-based Collaborative Filtering recommender model was implemented by Chandramouli et al. \cite{Chandramouli2011}. This approach is based on a stream processing system and focuses entirely on using explicit rating data. 

In our previous work \cite{lacic2017tailoring}, we presented a scalable recommender framework using a Microservices-based architecture to recommend a diverse set of entities in near real-time by leveraging the Apache Solr search engine.
In this work, we focus on adapting the non-probabilistic user-based Collaborative Filtering (UB-CF) algorithm \cite{schafer2007collaborative} 
to further improve its runtime performance by integrating a user pre-filtering step. 
This approach is especially useful in settings, in which it is not desirable to allocate additional resources but rather to optimize the usage of the available hardware.

\vspace{2mm}
\noindent \textbf{Bottleneck.}
Collaborative Filtering is usually accomplished in two steps: (i) the $k$-nearest neighbors are determined using a similarity metric (e.g., cosine similarity), and (ii) entities of these neighbors are recommended that the target user $u_{t}$ has not yet consumed \cite{schafer2007collaborative}. 
As  shown  in  previous  work  \cite{lacicscar},  both  steps  can  be  adapted  to search the data space in a scalable way and to retrieve the relevant content in near real-time. However, one performance bottleneck in this workflow is the number of neighbors that need to be processed (i.e., users which rate the same item as $u_{t}$).

While the neighborhood size $k$ is usually picked to be between 20 and 60 \cite{herlocker2002empirical}, it is still necessary to fetch the history of all neighbors and to calculate how similar a potential neighbor is to $u_{t}$. Moreover, the calculated similarities need to be sorted in order to pick the top-$k$ similar users. Common implementations of such operations have a complexity of $O(n \times log(n) )$\footnote{For example, Java's TreeMap implementation: \url{https://docs.oracle.com/javase/7/docs/api/java/util/TreeMap.html\#TreeMap(java.util.Map)}}. 
As such, the larger the neighborhood size of $u_{t}$ is, the larger the impact on the runtime performance could be.

\vspace{2mm}
\noindent \textbf{Contributions.}
In order to cope with such a performance bottleneck, in this paper, we present how to extend our scalable recommender system to save extra processing power by exploiting the Apache Solr search engine. We demonstrate that pre-filtering of users who exhibit a high number of overlapping entities can lead to better runtime performance as well as recommendation accuracy.

\section{Approach}
In this section, we present our approach for speeding up and enhancing CF-based recommendations with a user pre-filtering step.

\subsection{Adaptation of Collaborative Filtering with User Pre-Filtering}
\label{sec:adaptation}
In order to improve runtime, we could just decide to run the first step of CF in parallel, e.g., by having multiple processing nodes whose task is to fetch a user's history and calculate the similarity to the target user $u_{t}$. In this work, our aim is yet to increase the runtime performance in cases when there is also a limitation in terms of available processing resources.

Therefore, 
we propose to adapt the first step of UB-CF by pre-filtering the candidate set of possible similar users beforehand.  We do that in a greedy way by finding the top-$N$ candidate users with the highest overlap with respect to the available entity interactions (e.g., ratings). 
In this pre-filtering step, the similarity between $u_{t}$ and a possible candidate user $u_{c}$ is then calculated as follows: 
\begin{align}
OV(u_{t}, u_{c}) = |\Delta(u_{t}) \cap \Delta(u_{c})|
\end{align}
where $\Delta(u)$ corresponds to the set of entities some user $u$ has interacted with in the past.
As we will show next in Section \ref{sec:impl}, this can be done very efficiently by exploiting the Apache Solr search engine.
This way, we increase the probability that users with a high overlap will in the end be picked as the top-$k$ similar users. Also, by picking a reasonable value for $N$, we aim to positively influence the runtime performance of those users, which exhibit many neighbors.

\subsection{Implementation Details}
\label{sec:impl}
\begin{figure}[t!]
    \centering
    \includegraphics[width=0.49\textwidth]{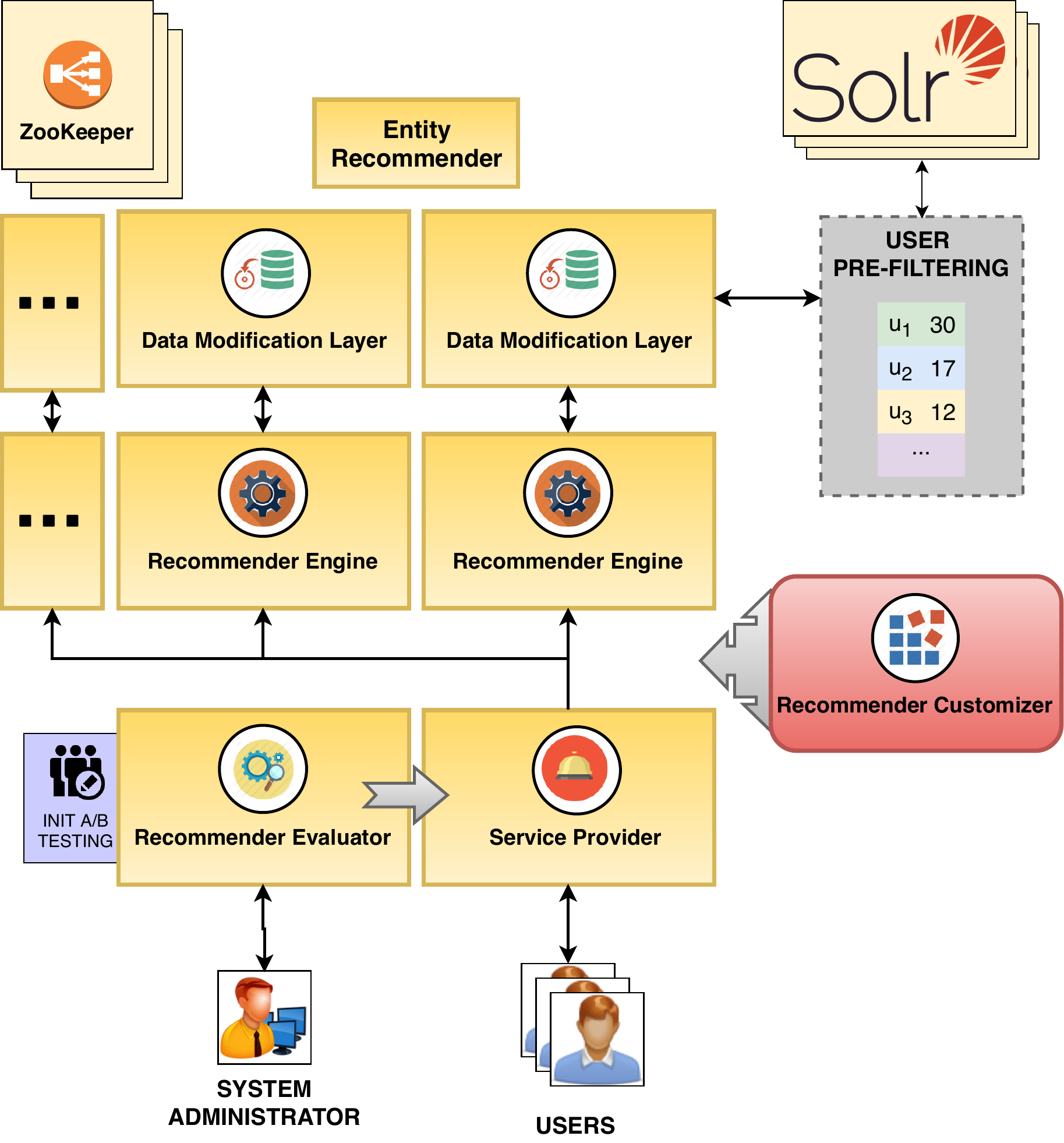}
    \caption{Architecture diagram of our scalable entity recommender framework that leverages the Apache Solr search engine to provide recommendations in near real-time. By pre-filtering users using Solr's facet functionality we not only receive the number of overlapping entites for the top-N users but also speed up the recommendation process.}
    \label{fig:scar}
    \vspace{-4mm}
\end{figure}

In our previous work \cite{lacicscar}, we have introduced a scalable software architecture, which can be applied to various entity recommendation scenarios. As seen in Figure \ref{fig:scar}, such an architecture allows us to recommend entities in an isolated environment. That is, every module can be deployed and started multiple times either on the same or on different machines and runtime performance can be guarantied by scaling individual nodes horizontally. To keep track of and coordinate all deployed nodes, we make use of Apache ZooKeeper \footnote{\url{http://zookeeper.apache.org/}}. An entity recommender is then set up of five modules which leverage Apache Solr to perform user pre-filtering in an efficient way.

\vspace{2mm}
\noindent \textbf{Service Provider} is the main entry point which acts as a proxy for the specific entity recommendation scenario (e.g., venues, movies, songs, etc.). It provides a REST-based interface to modules that are designated to handle the calculation of the requested entity recommendations as well as to store new data (e.g., interactions with the recommendable entities).

\vspace{2mm}
\noindent \textbf{Recommender Evaluator} aims to simulate user behaviour by splitting the data into training and test sets (see e.g., \cite{parra2010improving}). That is, for each user, a given number of entities is removed from the training set and added to the test set to be predicted. The difference between the recommended and the real data from the test set is then used to determine the success of the prediction. In addition to providing a diverse set of well-established recommender evaluation metrics, the evaluation procedure allows to simulate varying loads in order to better grasp the impact on the runtime when an increasing number of recommendations are requested. 

\vspace{2mm}
\noindent \textbf{Recommender Engine} contains recommender algorithms that use Apache Solr's efficient query language. In case of UB-CF, this allows us to immediately consider frequent data updates while providing real-time recommendations. Here, we calculate the probability that the target user $u_{t}$ will like an entity $e$ by the following formula \cite{schafer2007collaborative}:
\begin{align}
		pred(u_{t}, e) = \sum\limits_{u_{c} \in neighbors(u_{t}, e)}{sim(u_{t}, u_{c})}
\end{align}
where $neighbors(u_{t}, e)$ is the set of pre-filtered candidate neighbors of $u_{t}$ that have interacted with $e$ and $sim(u_{t}, u_{c})$ is the final similarity value between the users $u_{t}$ and $u_{c}$.

\vspace{2mm}
\noindent \textbf{Recommender Customizer}
allows to configure the implemented recommender approaches in form of recommender profiles. The sole purpose of these files is to customize a single recommender approach and to provide a reference to it. For instance, depending on the entity recommendation scenario that we wish to tackle, a Collaborative Filtering approach can be configured to use different kinds of similarity metrics (e.g., as shown in \cite{lacic15}), neighborhood sizes and use either explicit (e.g., ratings)  or implicit data (e.g., clicks) for recommending entities of similar users. 

\vspace{2mm}
\noindent \textbf{Data Modification Layer} acts as an agent between the recommender system and the Apache Solr search engine. By utilizing Solr,  we have the capability for horizontal scaling on the data storage side by creating either shards (i.e., splitting the data into smaller indices to increase the performance of search queries for huge data sets) or replicas (cloning the existing shards to another machine to increase the fault-tolerance of the whole system). 

\vspace{2mm}
\noindent \textbf{User Pre-Filtering} is performed very efficiently by exploiting Solr's facet functionality\footnote{\url{https://lucene.apache.org/solr/guide/7_0/faceting.html}}. This is basically an arrangement of search results into categories (e.g., user id's) along with numerical counts of matching documents. For example, if we perform a facet search on the user id over the whole document corpus (i.e., dataset) we would get for the top-$N$ users the exact count of corresponding entity interactions, where $N$ is a query parameter that needs to be provided to Solr and the resulting set is sorted in a descending order.

By defining a filter within the facet query to look only into entities from the target user's $u_{t}$ history, we not only reduce the search space but also get exactly the desired $OV(u_{t}, u_{c})$ values for the top-$N$ pre-filtered candidate users as defined in Section \ref{sec:adaptation}. Such a greedy pre-filtering of candidate users can be computed in milliseconds which in turns allows to speed up the generation of the final entity recommendations.

\begin{table}[t!]
\setlength{\tabcolsep}{4.5pt}
\centering
\begin{tabular}{c|c|cc|cc}
\specialrule{.2em}{.1em}{.1em}
    Data Type        & \multicolumn{1}{c|}{Density}  & \multicolumn{1}{c}{Mean}   & \multicolumn{1}{c|}{STD}    &  \multicolumn{1}{c}{Skewness} & \multicolumn{1}{c}{Kurtosis}	 \\\hline
    user - items  	 & 0.000015 & 2.32   &	54.08 & 227.53 & 62,884.43	 \\	
  \specialrule{.2em}{.1em}{.1em}
\end{tabular}
\caption{Statistics of our Foursquare dataset.}
 \label{tab:dataset_stats}
    \vspace{-6mm}
\end{table}

\section{Evaluation}
\label{sec:eval}
In this section, we describe our evaluation process, including our dataset, experimental setup as well as preliminary results.

\subsection{Dataset}
Traditionally, recommender systems deal with two types of entities, users and items. To show how  user pre-filtering can speed up and enhance recommendations, we used the Foursquare dataset provided by the authors of \cite{Levandoski2012, lars2014}. The dataset consists of 2,153,471 users, 1,143,092 venues (i.e., items) and 2,809,581 ratings. In order to make our dataset comparable, we present the summary of common statistical data measures for the user-item relationships in Table \ref{tab:dataset_stats}. The data density shows the proportion of actually known entities (e.g., ratings) to all possible entities that could be known by the user. 
This is rather a sparse dataset as the rating density is $0.000015$. 

Besides the mean entity assignments and their standard deviation, skewness and kurtosis \cite{Joanes1998} are also two important statistical measures. The skewness is a measure of the symmetry of a distribution. A symmetric distribution has a skewness of 0. In our case, the skewness is greater than 0 which means that the distribution is right-tailed (i.e., most data is concentrated on the left side of its function). Kurtosis is a measure of the distribution ``peakness'', where a higher kurtosis value signifies lower concentration around its mean. Especially in the case of the user - item  relationship, such a high kurtosis value means that the distribution has a sharper peak and broader tails.

\subsection{Experimental Setup}

We evaluated all users, which have at least one rated item in the training set. Thus, we extracted all users that
interacted with at least 11 items (= 58,046 users in total) and split the dataset in two different sets (training and test set) using a method similar to the one described in \cite{parra2010improving}. In other words, for each user, we withheld 10 items from the dataset and added them to the test set to be predicted. The rest of the data was used for training.

\vspace{2mm}
\noindent \textbf{Chosen Neighborhood Sizes and Recommendation Approaches.} 
With respect to neighborhood sizes, the distribution is right-tailed and the average neighborhood size is $764$, the median, however, is only $4$, while the maximum neighborhood size of a user is $125,046$. We determined values for $N$ in line with the literature \cite{herlocker2002empirical}, i.e., between 20 and 60. Specifically, we hypothesize that the same interval of values is valid for a greedy pick of candidate neighbors. We also evaluated $N=80$ and $N=100$ to test the impact of a larger candidate set on the accuracy. As shown in Table \ref{tab:approaches}, for each evaluated user seven recommendation approaches were evaluated.

\begin{table}[t!]
\setlength{\tabcolsep}{4.5pt}
\centering
\scalebox{0.93}{
\begin{tabular}{r|c}
\specialrule{.2em}{.1em}{.1em}
    \multicolumn{1}{c|}{Approach}      & \multicolumn{1}{c}{Description} 	 \\\hline
    MP  	 & A baseline Most Popular approach	 \\	
    $CF_{Full}$  	 & UB-CF which calculates similarities for all neighbors	 \\	
    $CF_{OV=20}$  	 & UB-CF with a greedy pick of top-20 overlapping users	 \\	
    $CF_{OV=40}$  	 & UB-CF with a greedy pick of top-40 overlapping users	 \\	
    $CF_{OV=60}$  	 & UB-CF with a greedy pick of top-60 overlapping users \\	
    $CF_{OV=80}$  	 & UB-CF with a greedy pick of top-80 overlapping users	 \\	
    $CF_{OV=100}$  	 & UB-CF with a greedy pick of top-100 overlapping users 	 \\	

  \specialrule{.2em}{.1em}{.1em}
\end{tabular}
}
\caption{Recommendation approaches evaluated in this paper.}
 \label{tab:approaches}
    \vspace{-6mm}
\end{table}

\vspace{2mm}
\noindent \textbf{Evaluation Metrics.} 
In our evaluation, we report the mean and standard deviation of the runtime performance as well as the recommendation accuracy in terms of Precision (P), Recall (R), Normalized Discounted Cumulative Gain (nDCG) and User Coverage (UC) \cite{Shani2011}.

\begin{table*}[t!]
\setlength{\tabcolsep}{10pt}
\centering
\scalebox{0.99}{
\begin{tabular}{c|c|c|c|c|c|c|c}
\specialrule{.2em}{.1em}{.1em}
  \multicolumn{1}{c}{} & Approach & $\overline{T}$ (ms) & $\sigma$ (ms) & $P@10$ & $R@10$ & $nDCG@10$ & $UC$   \\\hline \hline
  \multicolumn{1}{c}{} & Most Popular  		& 78.59 & 20.00	 & $.0285$ & $.0285$ & $.0232$	& $\textbf{100\%}$ \\			
  \hline 
  \multirow{6}{5pt}{\centering{\begin{sideways}\centering{CF}\end{sideways}}} &
     $CF_{Full}$		   & 2,053.45 & 9,600.63  	& $.0611$ ($.0918$) & $.0527$ ($.0792$) & $.0316$ ($.0475$) & $66.56\%$	 \\	\cline{2-8}
  &  $CF_{OV=20}$  & \textbf{59.56} & \textbf{60.08}          & $.0586$ ($.0890$) & $.0541$ ($.0821$) & $.0318$ ($.0483$) & $65.87\%$		 \\	
  &  $CF_{OV=40}$ & 65.47 & 69.61          & $.0689$ ($.1042$) & $.0645$ ($.0974$) & $.0378$ ($.0571$) & $66.21\%$		 \\	
  &  $CF_{OV=60}$ & 74.62 & 85.83          & $\textbf{.0724}$ ($\textbf{.1095}$) & $\textbf{.0678}$ ($\textbf{.1026}$) & $\textbf{.0396}$ ($\textbf{.0599}$) & $66.10\%$		 \\	
  &  $CF_{OV=80}$  & 82.40 & 102.75         & $.0707$ ($.1077$) & $.0661$ ($.1007$) & $.0386$ ($.0588$) & $65.62\%$		 \\	
  &  $CF_{OV=100}$ & 87.38 & 115.17 	    & $.0693$ ($.1055$) & $.0646$ ($.0983$) & $.0373$ ($.0568$) & $65.70\%$ \\ 
  \specialrule{.2em}{.1em}{.1em}
\end{tabular}
}
\caption{For all recommendation approaches, we report the mean runtime, as well as the accuracy and user coverage (UC). The values in brackets represent the results normalized to the UC in the row, while the ones without are calculated on a 100\% UC (i.e., all 58,046 users). Our neighborhood pre-filtering approach leads to improvents with respect to runtime and accuracy.}
 \label{table:result}
 \vspace{-4mm}
\end{table*}

\subsection{Preliminary Results}
Our evaluation results are summarized in Table \ref{table:result}. The experiments have been executed on an IBM System x3550 server with two 2.0 GHz six-core  Intel  Xeon  E5-2620  processors, a  1TB  ServeRAID  M1115 SCSI Disk and 128 GB of RAM using one instance and Ubuntu 14.04.1. with Apache Solr 4.10.2.

All performance metrics are reported for $10$ recommended items (k=$10$).
On average, $CF_{Full}$ took approximately $2$ seconds with a rather high standard deviation of $9.6$ seconds. With the adapted CF approaches, where we calculated the similarity only on the top-$N$ overlapping users, the runtime performance drastically improves (i.e., below $90$ ms, which is more than $23$ times faster than the  $CF_{Full}$). Such a runtime is even comparable to the one of the simple MostPopular (MP) baseline.

\begin{figure}[t!]
    \centering
    \includegraphics[width=0.49\textwidth]{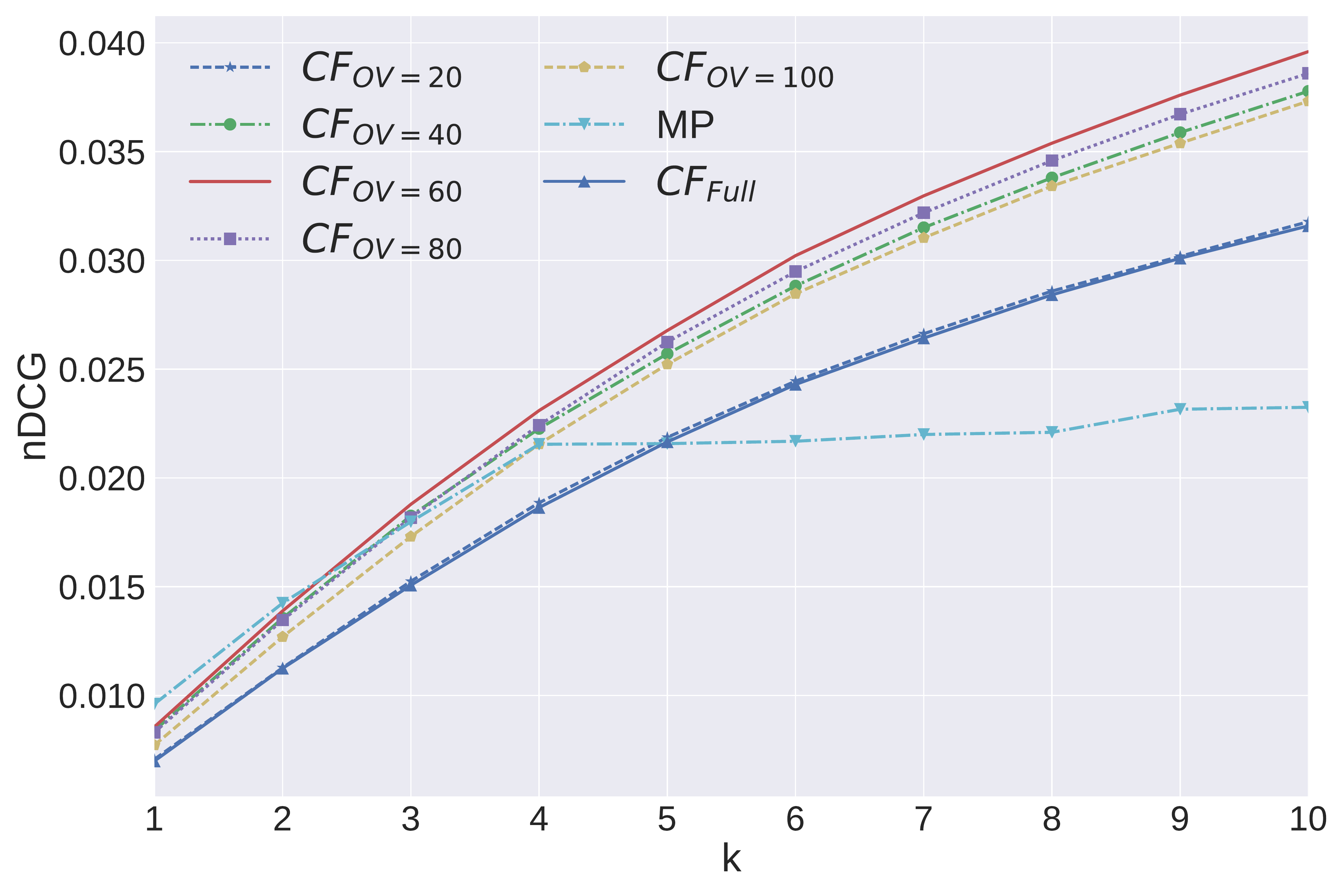}
    \caption{nDCG  plot showing the difference in accuracy when using top-$N$ overlapping users as candidate similar users. }
    \label{fig:ndcg}
    \vspace{-4mm}
\end{figure}

Interestingly, the accuracy also increases when we pre-filter the candidate set of similar users, as also shown in terms of nDCG in Figure \ref{fig:ndcg} for different values of $k$. We achieved the best performance, both in terms of runtime and accuracy, by utilizing the top-$60$ overlapping users. Here, the runtime performance was almost the same as when running the MP baseline. 
This suggests that creating a pre-filtered candidate set of similar users not only yields better runtime performance but can also contributes to a higher recommendation accuracy.

\section{Conclusion and Future Work}
\label{sec:conc}
In this paper, we presented work-in-progress on adapting Collaborative Filtering by integrating a user pre-filtering step to speed up and enhance entity recommendations.
Specifically, we adapt the approach by applying user pre-filtering, in which we generate a smaller set of candidate neighbors in a greedy fashion (i.e., by focusing on neighbors with a higher number of overlapping entities). 
Our results suggest that our pre-filtering approach can not only achieve a better runtime performance but also is able to increase the overall accuracy compared to a classic CF algorithm without user pre-filtering.

\vspace{2mm}
\noindent \textbf{Limitations and Future Work.}
One limitation of our work is that we evaluated our approach only on one dataset.
As a next step, we want to validate our results in a more comprehensive study using datasets with different types of entities that can be recommended. Here, we especially aim to validate our approach in course of the Analytics for Everyday Learning (AFEL) project\footnote{\url{http://afel-project.eu/}} \cite{d2018afel} for recommending learning resources. This would also allow us to evaluate this approach in course of an online study to measure the real user acceptance of the recommendations.

\vspace{2mm}
\noindent \textbf{Acknowledgments.}
The authors would like to thank the Social Computing research area of the Know-Center GmbH and the AFEL consortium for their support. This work was funded by the Know-Center GmbH Graz (Austrian FFG COMET Program) and the European-funded H2020 project AFEL (GA: 687916).

\bibliographystyle{abbrv}

\end{document}